# Influence of the Substrate Material on the Optical Properties of Tungsten Diselenide Monolayers

Sina Lippert[1,+], Lorenz Maximilian Schneider[1,+], Dylan Renaud[1], Kyung Nam Kang[2], Obafunso Ajayi[3], Marc-Uwe Halbich[1], Oday M. Abdulmunem[1], Xing Lin[1], Jan Kuhnert[1], Khaleel Hassoon[1], Saeideh Edalati-Boostan[1], Young Duck Kim[3], Wolfram Heimbrodt[1], Eui-Hyeok Yang[2], James C. Hone[3], and Arash Rahimi-Iman[1,*]

1) Department of Physics and Materials Sciences Center, Philipps-Universität Marburg, Marburg, 35032, Germany

2) Department of Mechanical Engineering, Stevens Institute of Technology, Hoboken, New Jersey, 07030, USA.

3) Department of Mechanical Engineering, Columbia University, New York, New York 10027, USA

[+] these authors have contributed equally

[*] a.r-i@physik.uni-marburg.de

**Abstract:**

Monolayers of transition-metal dichalcogenides such as $WSe_2$ have become increasingly attractive due to their potential in electrical and optical applications. Because the properties of these 2D systems are known to be affected by their surroundings, we report how the choice of the substrate material affects the optical properties of monolayer $WSe_2$. To accomplish this study, pump-density-dependent micro-photoluminescence measurements are performed with time-integrating and time-resolving acquisition techniques. Spectral information and power-dependent mode intensities are compared at 290K and 10K for exfoliated $WSe_2$ on $SiO_2/Si$, sapphire ($Al_2O_3$), $hBN/Si_3N_4/Si$, and $MgF_2$, indicating substrate-dependent appearance and strength of exciton, trion, and biexciton modes. Additionally, one CVD-grown $WSe_2$ monolayer on sapphire is included in this study for direct comparison with its exfoliated counterpart. Time-resolved micro-photoluminescence shows how radiative decay times strongly differ for different substrate materials. Our data indicates exciton-exciton annihilation as a shortening mechanism at room temperature, and subtle trends in the decay rates in correlation to the dielectric environment at cryogenic temperatures. On the measureable time scales, trends are also related to the extent of the respective 2D-excitonic modes' appearance. This result highlights the importance of further detailed characterization of exciton features in 2D materials, particularly with respect to the choice of substrate.

# Introduction

2D materials such as $MoS_2$, $WS_2$ and $WSe_2$ belong to the family of transition metal dichalcogenides (TMDs) which have recently attracted a vast amount of attention for their remarkable and unusual properties. As a semiconducting alternative to graphene, TMDs have promising applications in photonics [1, 2], optoelectronics [3, 4], valleytronics [5], field effect transistors [6], gas sensors [7], mechanical resonators [8,9] and energy storage devices [10].

In 1970, Consadori and Frindt produced bilayer $WSe_2$ for the first time by mechanical exfoliation [11]. Today, the "scotch tape method" is the most used method [12-16] to prepare monolayers (MLs) of $WSe_2$ from its bulk counterpart. However, $WSe_2$ layers have been also fabricated using chemical exfoliation [17-19], chemical vapor deposition (CVD) [20-24], metal-organic chemical vapor deposition (MOCVD) [25], hydrothermal exfoliation [26], liquid exfoliation [27-29], and physical vapor deposition [30-31]. Due to the existence of these various fabrication techniques, the focus has now shifted to the production of high-quality MLs [16, 32, 33].

Reflection contrast [34-36], transient absorption [35], time-integrated photoluminescence (PL) [37, 38] and time-resolved photoluminescence (TRPL) [39, 40] experiments, have been performed to study the emission properties of $WSe_2$. In prior studies on $WSe_2$, layers were deposited on $SiO_2$/Si substrates [41, 42], sapphire [43, 44], graphene [45], and fused silica (quartz) [34, 46] or sandwiched between layers of hBN [47]. Based on the literature on $WSe_2$ and its properties, it has been identified as an ideal/suitable testbed for investigating the impact of substrate properties on its excitonic species. In addition, $WSe_2$ possesses good luminescence at room temperature (RT) and reasonable emission at low temperatures (LT). Finally, prior studies have confirmed the existence of excitons, trions [34, 38, 48] and more recently even biexcitons [49] and dark excitons [50]. Ultimately, the role of the dielectric environment and surface properties on excitons shall be unravelled in the future. Nevertheless, first studies indicate resonance shifts due to surface quality [51], strain and tensions in the material [38, 52, 53], and water moisture [54]. Others indicate that the excitonic resonance remains fixed even though the dielectric environment is altered [55]. This process is understood as an expected change of binding energy being compensated by a simultaneous bandgap renormalization [56] which can take place in $WSe_2$ [33, 57].

To date, a decisive comparison of ML samples showing the effect of the substrate material on a single type of ML material's emission signatures, on its time-dependent emission characteristics and on its Raman spectra has not yet been performed. Herein, we investigate time-integrated μPL spectra and corresponding time-resolved emission of ML $WSe_2$ (as a representative of this material class) using exfoliated $WSe_2$ isolated on $SiO_2$, sapphire, $MgF_2$ and hBN/$Si_3N_4$, together with a CVD-grown $WSe_2$ monolayer on sapphire, in order to show strong similarities and distinct differences in the emission pattern based on the substrate-material choice.

# Experiment

## Monolayer samples

Mechanically exfoliated WSe$_2$ MLs have been transferred onto n-type SiO$_2$(300nm)/Si, sapphire, multilayer-hBN (>10nm) on Si$_3$N$_4$(75nm)/Si and MgF$_2$ substrates and studied as prepared (for details on the sample fabrication see methods section). Although suggestions were made in the literature for the improvement of optical properties for ML materials by chemical treatment [58] and for mechanically stacked systems [59], a decisive comparison requires untreated samples to be investigated, which were fabricated under similar circumstances using the same procedures. Additionally, WSe$_2$ growth was conducted via a low pressure CVD process on a sapphire substrate. The detailed conditions of growth are similar to the growth condition of WS$_2$ reported elsewhere [60, 61]. This growth technique produced both millimeter-sized polycrystalline WSe$_2$ MLs and single crystalline WSe$_2$.

**Table 1 | Sample list**

| Substrate/Monolayer Material | Preparation Method | Refractive Index of Substrate at 750 nm | Reference (www.refractiveindex.info) |
|---|---|---|---|
| SiO$_2$/ WSe$_2$ | Exfoliated | 1.474 | [62] Gao et al (2013) |
| Sapphire/ WSe$_2$ | Exfoliated | 1.768 | [63] Malitson and Doge (1972) |
| Sapphire/ WSe$_2$ | CVD | 1.768 | [63] Malitson and Doge (1972) |
| MgF$_2$/ WSe$_2$ | Exfoliated | 1.377 | [64] M. J. Dodge (1984) |
| Si$_3$N$_4$/hBN/ WSe$_2$ | Exfoliated | 2.017(Si$_3$N$_4$) / 2.200(hBN) | [65] Philipp (1973) / [66] Gielisse et al. (1967) |

Here, the choice of substrates has been made for various reasons. Most importantly, a comparison of common transparent and opaque materials is desired. The chosen materials all exhibit a different refractive index, ranging from 1.38 to 2.2 (see Tab. 1, and Tab. SI.2 in the supporting information for further details). While oxidized Si (SiO$_2$ on Si) has become the standard platform for the investigation of 2D materials, other materials such as sapphire (Al$_2$O$_3$) and MgF$_2$ are becoming increasingly attractive due to their transparency. Combined with large area ML coverage by epitaxy (such as CVD), transparent materials can enhance the applicability of monolayer materials in optical devices while simultaneously giving access to experiments which require a transmission geometry. Taking into account the recent hunt for alternatives to SiO$_2$, Si$_3$N$_4$ has also been considered. It was recently introduced as a substrate material with improved optical contrast when used as a sub-100 nm layer on Si [67]. However, due to the potential of multi-layer hBN as an atomically smooth buffer layer [59], Si$_3$N$_4$ has been covered with exfoliated hBN to restore WSe$_2$'s optical properties by preventing ML corrugation as a result of substrate surface roughness.

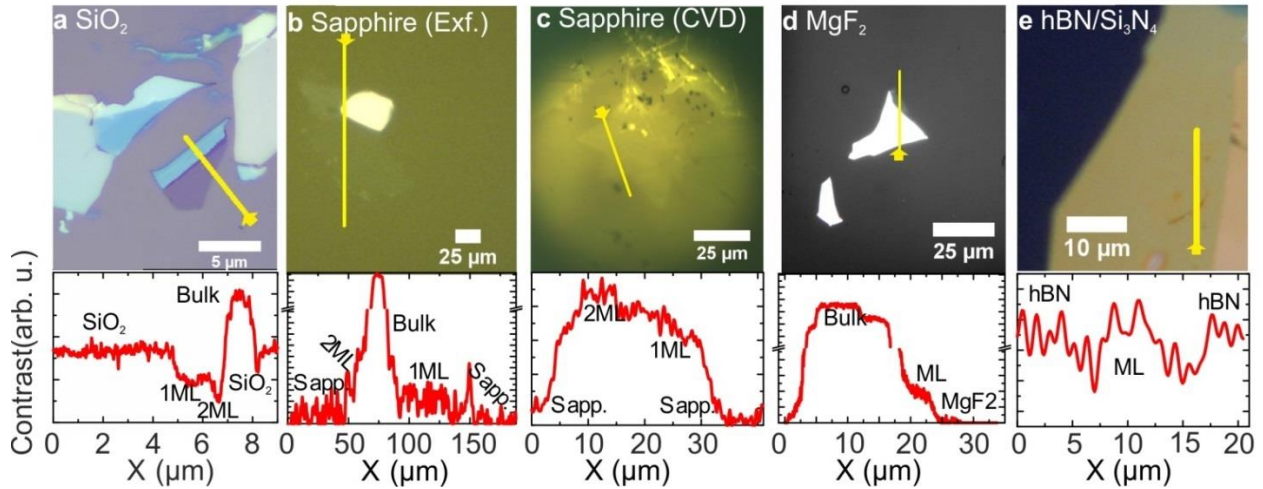

**Figure 1 | Microscopic images of the measured WSe$_2$ monolayers deposited on different substrates with cross sections.** The substrates are (a) silicon dioxide, (b) sapphire exfoliated (exf.), (c) sapphire (CVD), (d) MgF$_2$, while in (e), a large flake of multilayer hBN is located underneath the WSe$_2$. Excluding (c), the investigated samples have been fabricated by mechanical exfoliation. The yellow line in each image indicates the path, along which the brightness cross-sections are taken and which are displayed in the respective charts in the lower row. Given the contrast variations, the layers can be identified and are marked in the cross sections.

Microscopic pictures of the measured WSe$_2$ ML flakes on different substrates are shown in Figure 1. The micrograph in Figure 1a was recorded using a 100x magnification objective, while b through e were recorded with a 20x objective. In order to characterize the layer numbers, the optical contrast was evaluated along the path depicted in the figure as a yellow line. For this, the open-source software ImageJ was used [68]. The resulting cross sections are shown underneath the respective micrographs in Figure 1. The steps are clearly visible in the cross section and were used to identify the monolayer sections [68]. The respective ML sections have been marked together with bulk and hBN sections (see Fig. 1). WSe$_2$ MLs have also been verified by Raman spectroscopy and µPL, which show the Raman signature and spectral features of ML WSe$_2$ for all investigated samples (for Raman data, see Fig. SI.1 in the supporting information).

**Experimental setup**

The micro-photoluminescence (µPL) and time-resolved (TR) µPL (in the following simply referred to as TRPL) measurements were performed with a pulsed Titanium-Sapphire (Ti:Sa) laser with a tuneable emission wavelength of 700-1000 nm, a pulse duration of 100 fs and a repetition rate of 80 MHz. The light from the laser was frequency doubled by nonlinear optics to provide an excitation wavelength of 445 nm. A schematic diagram of our optical setup to perform µPL measurements is shown in Fig. 2.

The samples were mounted onto the cold-finger of a continuous flow cryostat, where temperature could be varied between 10 K and 290 K using a cooling system with liquid helium. The power-dependent µPL and TRPL measurements were conducted at 10 K and at 290 K under ultra-high vacuum conditions. During TRPL and µPL measurements, the laser spot size on the samples was approximately 4 μm. The time-averaged excitation densities at the pump spot delivered by the pulsed laser were determined to be 340, 1000, 2900, and 3400 W/cm².

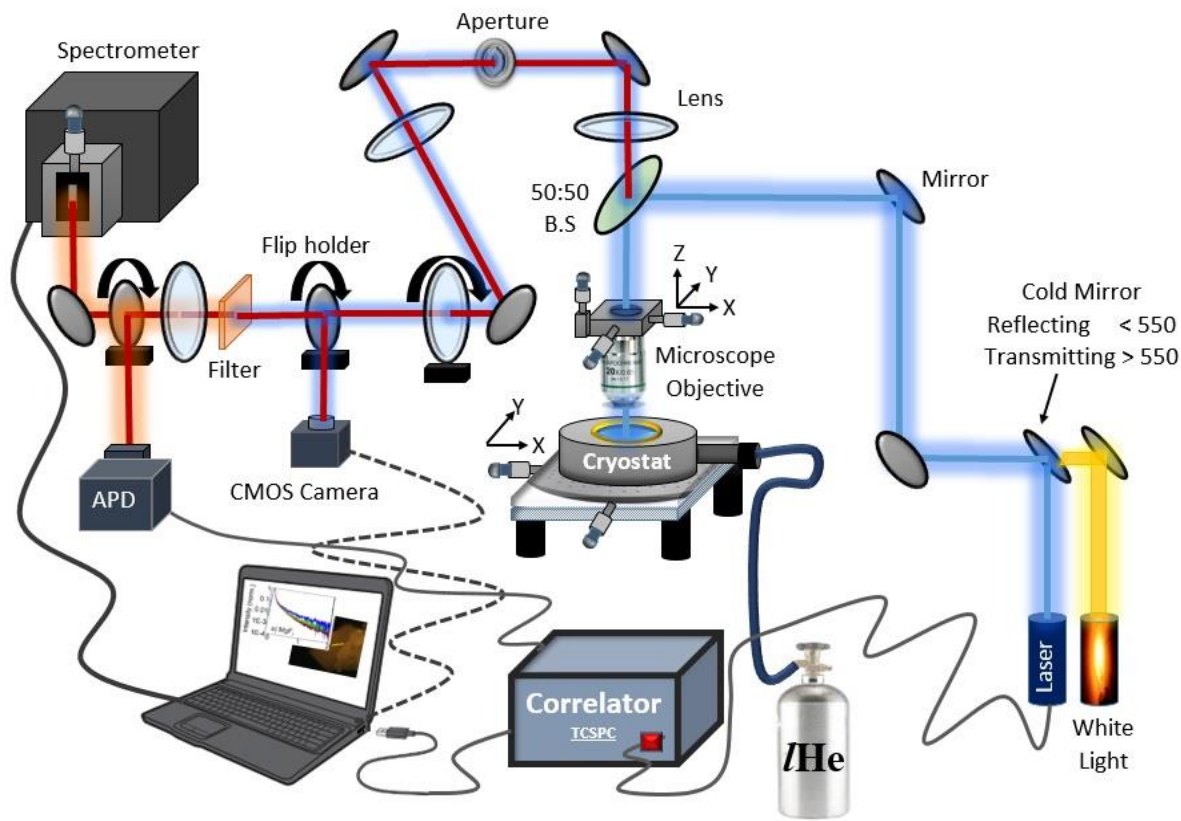

**Figure 2 | Schematic diagram of the micro-photoluminescence setup.** The light of the Ti:Sa laser is focused onto the sample using a conventional confocal microscope setup. The setup uses a 20x objective to focus the light onto a 4 μm spot on the sample. The light is collected through the same objective. An iris aperture is used in the sample projection plane for spatial selection of the detection area. The sample can then be imaged using a removable lens and flip mirror in conjunction with a CMOS camera. The light can be focused onto the spectrometer slit for the acquisition of time-integrated spectra with an ICCD or onto an APD for transient µPL measurements, both at room temperature and at 10 K.

The laser beam was focused under normal incidence onto the sample using a 20x microscope objective (NA 0.42). µPL emission from the sample was collected by the same objective. For spatial selection, an iris aperture in the real-space projection plane was used. In order to image samples, a CMOS camera in combination with optical lenses and mirrors is included. The µPL is collected by a grating spectrograph using a grating with 300 grooves/mm and an air-cooled intensified CCD, whereas for TRPL, an avalanche photo-diode (APD) is used with a time-correlated single-photon counting (TCSPC) unit. For TRPL measurements, the complete ML signal is acquired by the APD spectrally integrated behind a long-pass filter. For more details and explanations of the experimental methods, please see the methods section and supporting information.

**Results and Discussion**

**Photoluminescence Spectra**

Time-integrated µPL spectra at 290K and 10K have been acquired and are presented in Fig. 3. To investigate the power dependence of the µPL signal, four different excitation powers have been used. The excitation powers measured and verified before the beam splitter are 87 µW (blue), 250 µW (green), 720 µW (red) and 870 µW (black), corresponding to at maximum 43 µW, 125 µW, 360 µW and 430 µW after the beam splitter, which gives estimated mean pump densities of 340, 1000, 2900, and 3400 W/cm², respectively. To discriminate the contributing spectral components of the PL signal, a multi-peak evaluation with Gaussian peaks was performed. The sum of the Gaussian peaks is shown in light grey (which can be hardly distinguished from the emission spectra owing to the strong matching), while the single Gaussian peaks are shown in dark grey. The corresponding fit parameters *energy* (Peak position) and *line width (FWHM)* are summarized in the supporting information (see Fig. SI.3 and SI.4), while the integrated peak intensities are discussed below (see Fig. 4).

At room temperature (RT), a main peak and a red-shifted shoulder were observed for all substrates. Interestingly, the main peak attributed to the RT exciton is found at 1.66 eV for all substrates except for CVD-grown WSe$_2$ on sapphire (1.63 eV). The obtained excitonic energies for exfoliated sample are well comparable to Godde et al. [69]. Nevertheless, the position of the exciton for the CVD-grown ML on sapphire is significantly different from that of bilayer emission and it agrees well with the result presented by Huang et al. [22]. A spectral comparison of ML and bilayer emission is shown in the supporting information (see Fig. SI.2). Here, the particularly broad peak can be an indicator for a superposition of exciton and trion peaks, but can also hint at the mere occurrence of trions owing to a possibly larger rate of defect states (lattice dislocations, donor/acceptor states) as a possible consequence of CVD growth. Alternatively, one can understand the red shift of the exciton mode as a strain-induced effect owing to the hot temperatures in the furnace at which CVD growth takes place, while ML samples prepared by exfoliation at room temperature have very similar emission energies to one another. Since CVD ML are formed on the hot surface out of the vapor phase at elevated temperatures, the consecutive cooling after growth leads to tensions caused by the mismatch of the thermal expansion coefficients (TECs) of WSe$_2$ (in plane coefficient) TEC=(1.1x10$^{-5}$/K~1.4x10$^{-5}$/K) [70,71] and sapphire TEC=(5x10$^{-6}$/K~8.3x10$^{-6}$/K) [72] at the relevant temperatures (300-900K), which differ by a factor of 1.5~2. Qualitatively, the TEC of sapphire is less than that of WSe$_2$ over this temperature range. This leads to a scenario in which WSe$_2$ wants to shrink faster than the substrate during the cooling process but is stretched due to tension as a result of surface adhesion. Such conditions can indeed affect excitonic modes [38], as a stretched lattice with increased mean particle distances in the plane can exhibit a band gap energy reduction. This phenomenon has been confirmed during a high-temperature optical spectroscopy of ML WS$_2$ [73].

Similar to Ref. [74], we attribute the shoulder in the µPL spectra of our samples to trion emission. The peak center energies obtained from multi-peak fitting representing main and shoulder peaks are summarized in Tab. 2 (see below). The data shown correspond to the averaged energies obtained for different excitation densities. Nevertheless, no significant peak shift has been found depending on the pump power. The errors correspond to the standard derivation of the values obtained at the four powers.

Interestingly, a subtle correlation between the refractive index of the substrate and the energetic position of the RT exciton modes can be observed (see Tab. 2), although the refractive index was only changed for the half space, i.e. on one side of the sample. However, the details of such dependency cannot be clarified within the scope of this work. Nevertheless, it is expected that the influence of a dielectric medium change for MLs on the optical dipoles of MLs is noticeable even when one half space remains at n=1, since the field lines of such Rydberg-like dipoles penetrate into the MLs surrounding environment (as described by the literature, cf. [75]). Here, the lowest refractive index material, $MgF_2$, has the lowest exciton energy while the highest refractive index substrate, $hBN/Si_3N_4$, features the highest, with a total difference of about 9 meV in peak positions (see Tab. 1). Nevertheless, there is no unambiguous correlation with the refractive index of the substrate material. For example, this correlation has not been found for the emission attributed to the trion, and also not for the features measured at 10 K. The observation of a trend at room temperature also excludes the CVD-grown ML on sapphire because of its strong peak shift. This deviation of the CVD result at room temperature can be explained by the ML fabrication technique; CVD ML fabrication takes place at high temperatures and can introduce strain to both the substrate surface and the deposited material as a result of the annealing process and subsequent cooling. Moreover, it is expected that the incorporation of defects and impurities into the 2D lattice is stronger for CVD grown MLs than for exfoliated crystals, suggesting a broader spectral distribution of RT emission and more emission from defects for CVD MLs.

In sum, it seems that the substrate only slightly affects the µPL features of $WSe_2$ at room temperature (cf. fit parameters summarized in the supporting information, Figs. SI.3 and SI.4). This may increase the importance of $WSe_2$ as a 2D material due to its ability to be deposited on a variety of substrates without losing its general spectroscopic attributes.

In contrast to the RT case, at T=10K four different PL emission features were identified for ML $WSe_2$ isolated on $SiO_2$ and sapphire (CVD), whereas only three peaks were observed for $WSe_2$ exfoliated on sapphire (exf.), $hBN/Si_3N_4$ and $MgF_2$ substrates. The four peaks can be identified as exciton (1.73 eV), trion (1.71 eV), biexciton (1.69 eV) and localised states (1.66 eV and below) and show good agreement with the energetic positions found in [38, 49, 69]. Table 2 summarizes the results of Fig. 3 (a-j).

In general, the energetic positions of all features are comparable within the fit accuracy for all substrates except for the CVD grown sample. Therefore no trend can be seen as a function of the refractive index. For some samples, no distinct exciton and trion features were obtained, and a significant separation of such two species was not found. Consequently, the higher energy shoulder peak(s) (even if two species could be presumed) was fitted with one Gaussian peak. The exfoliated MLs on sapphire and $MgF_2$ clearly show broader central peaks (FWHM~25 to 40 meV) compared to the other samples (FWHM~15 meV) (for more details see the supporting information). This could be an indicator of different surface qualities or differences with respect to defect states and impurities. Nevertheless, the values obtained for excitonic features are in good agreement with those reported by Wang et al. [38] (exciton: 10 meV, trion: 15 meV) and exhibit quite narrow line widths and comparable energy positions within the batch of different samples. Here, spectral similarities are very pronounced, while no trend in relation to the refractive index is evidenced. However, this can be understood as many factors can influence the spectral properties, such as strain effects in low-temperature ML-substrate compounds and the compensation of opposing effects such as binding energy modifications and gap renormalization, which cannot be quantified readily in such a study.

**Relationship between laser power and µPL intensity**

To get further insight into the recombination processes, double-logarithmic plots of the power dependence ($P_L$) of each peak's intensity ($I_{PL}$) are presented in Fig. 4. This type of analysis technique is useful since the emission of localized states [69] or bound states [49] grow more slowly than exciton emission with increasing pump power and exhibit sub-linear power dependence. Consequently, using $P_L$-$I_{PL}$ measurements, excitonic features can be identified by their super-linear behavior [49].

All the logarithmic plots in Fig. 3 (a-j) can be described by the power law-equation, i.e. $I_{PL} \propto P_L^{\alpha}$ [49], where α is the linearity or exponent factor. The corresponding peaks' center energies resulting from the fits are given in the legends and are summarised in the supporting information. The up to four data series (solid symbols) correspond to the up to four distinct features attributed to the exciton (black square), trion (red circle), biexciton (green up-triangle) and localized state (blue down-triangle) in the µPL spectra of the different samples.

For the room temperature measurements of $WSe_2$ on $SiO_2$, $WSe_2$ exfoliated on sapphire and $WSe_2$ on $hBN/Si_3N_4$, we found values for alpha close to 0.5. While for CVD-grown $WSe_2$ on sapphire and $MgF_2$, another behavior is observed. Referring to the rate equations describing the recombination of free and localized carriers, 0.5 corresponds to the recombination of electron-hole pairs at localized centers [76]. The slightly larger values for CVD-grown $WSe_2$ on sapphire can be explained by the presence of defects [74]. For $MgF_2$ it seems that for higher powers, the excitonic emission vanishes while the trion emission increases.

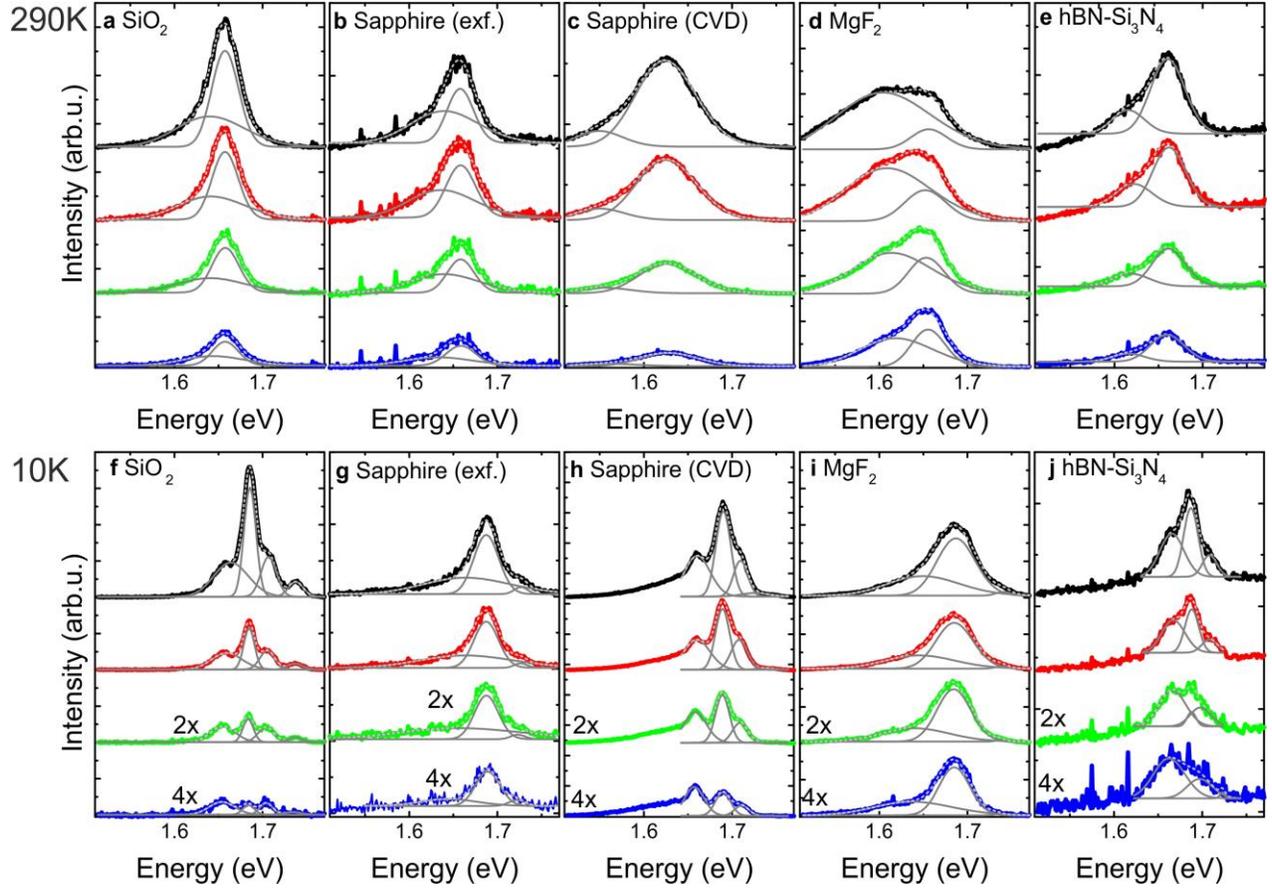

**Figure 3 | Time-integrated PL spectra of WSe$_2$ monolayers on different substrates at 290K and 10K for different excitation powers**. The excitation densities shown are 340 (blue), 1000 (green), 2900 (red), and 3400 W/cm² (black). The spectra have been fitted each with the sum of (at least) two Gaussian peaks (light grey dashed lines), with the underlying Gaussian peaks shown in dark grey. For better visibility, the spectra at 10K for the lowest two powers have been scaled.

For low temperature measurements, the observed linearity factors show a different behavior in comparison to the ones obtained at room temperature. The α -value for the exciton lies between 1.1 and 1.3 which is comparable to the values given in Ref. [74] and within the theoretical expectation given by Ref. [76]. For the trion, we obtain values of 1.1. These are comparable to earlier reported values from Yan et al. [77]. A possible reason for the smaller α -value of the trion for WSe$_2$ on hBN/Si$_3$N$_4$ might be related to the obtained background PL signal of Si$_3$N$_4$ which reduced the quality of the peak fitting at low excitation densities. Indeed, at higher pump densities, the slope recovers from the negative effect of background PL. The values of the superlinearity on SiO$_2$/Si, Sapphire (CVD) and hBN/Si$_3$N$_4$ for the biexciton (1.2 to 1.5) match the expected higher value of 1.5 reported earlier [49]. Although the other two samples exhibit an emission at the same energetic position as the biexciton, the value for α and the line width of the peak do not match the expected value. As discussed earlier, these samples probably inherit more defects leading to more localized emission. Unsurprisingly, the localized states show an α-value strongly below 1 and match the expected value of 0.5 [69]. A summary of all experimentally determined α factors is given in Tab. 2.

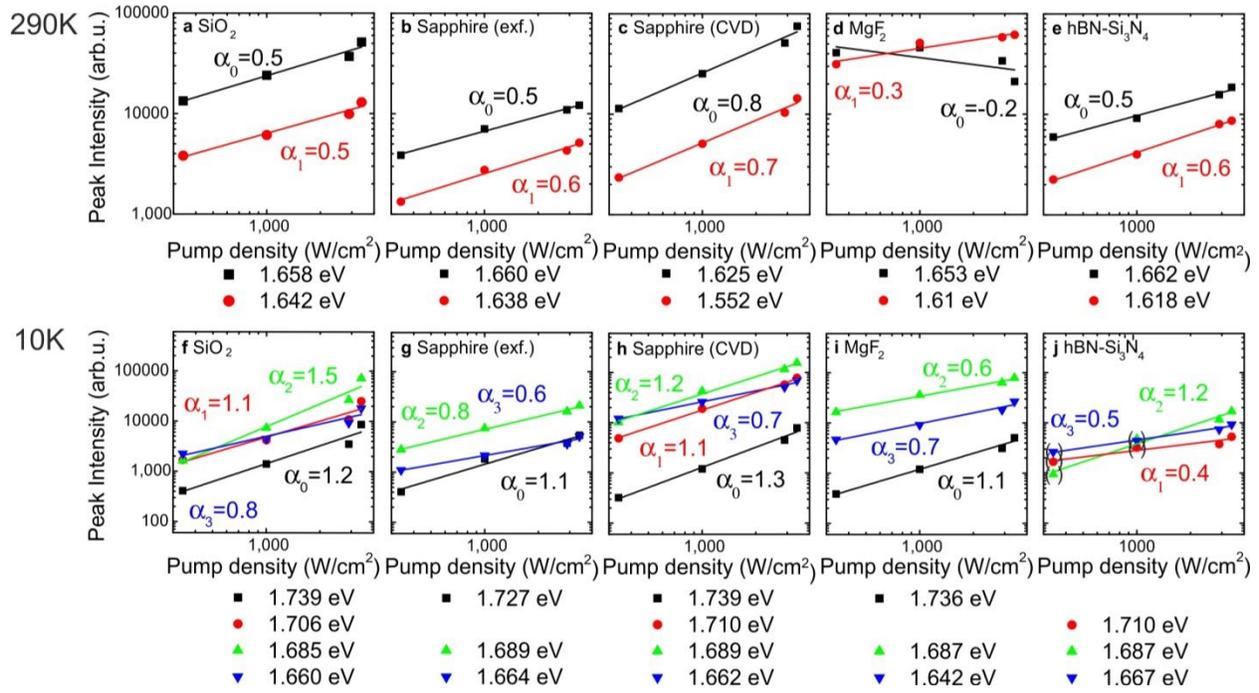

**Figure 4 | μPL intensity as a function of laser power for different substrates at 290K and 10K.** The power dependence has been plotted on a double-logarithmic scale. The data series correspond to the Gaussian peaks used to fit the PL-spectra. The average peak energies resulting from multi-peak fitting is given in the legend. Furthermore, the series have been fitted to a power law (line) to get further insight into the recombination mechanism and for the sake of comparison. The four data series (symbols) summarized correspond to the features attributed to the exciton (black square), trion (red circle), biexciton (green up-triangle) and localized state (blue down-triangle) in the μPL spectra.

**Table 2 | Calculated energy values for exciton, trion, biexciton and bound states at 10K and 290K in $WSe_2$ MLs deposited on different substrates**. Energies are rounded to three decimal figures for the sake of legibility. The provided errors with three decimal figures correspond to the standard deviation of the values obtained for the four pump-power settings. The values of the exponent factor are also indicated in the respective row and column. Results shown in parentheses cannot be attributed to the respective species.

| Temp. (K) | Feature | $SiO_2$/Si | Sapphire | Sapphire (CVD) | $MgF_2$ | hBN/$Si_3N_4$ |
|---|---|---|---|---|---|---|
| 290 | Exciton (eV)/ | 1.658 ±0.000 | 1.660 ±0.000 | 1.625 ±0.002 | 1.653 ±0.002 | 1.662 ±0.001 |
|  | α -value | 0.5 | 0.5 | 0.8 | -0.2 | 0.5 |
| 290 | Trion (eV)/ | 1.642 ±0.002 | 1.636 ±0.003 | 1.552 ±0.004 | 1.610 ±0.045 | 1.618 ±0.005 |
|  | α -value | 0.5 | 0.6 | 0.7 | 0.3 | 0.6 |
| 10 | Exciton (eV)/ | 1.739 ±0.002 | 1.727 ±0.002 | 1.739 ±0.013 | 1.736 ±0.003 |  |
|  | α -value | 1.2 | 1.1 | 1.3 | 1.1 |  |
| 10 | Trion (eV) | 1.706 ±0.001 |  | 1.710 ±0.001 |  | 1.705 ±0.010 |
|  | α -value | 1.1 |  | 1.1 |  | 0.4 |
| 10 | Biexciton (eV)/ | 1.685 ±0.001 | (1.689 ±0.001) | 1.689 ±0.001 | (1.687 ±0.002) | 1.687 ±0.005 |
|  | α -value | 1.5 | (0.8) | 1.2 | (0.6) | 1.2 |
| 10 | Localised state (eV)/ | 1.660 ±0.005 | 1.664 ±0.016 | 1.662 ±0.004 | 1.642 ±0.005 | 1.667 ±0.002 |
|  | α -value | 0.8 | 0.6 | 0.7 | 0.7 | 0.5 |

**Time Resolved Photoluminescence (TRPL)**

In Figure 5 a bar chart of the different decay times for all five substrates is shown. Triexponential fits were used to systematically extract the first (fast) and second (slow) decay time from each curve (an example is shown as an inset in Fig. 5 and others can be found in the supporting information, together with the fit equation and an overview on the fit parameters, including first, second and third time constants and their amplitudes, respectively). The triexponential fits were used for all data; although for some data biexponential fits would have been enough (that would lead the extracted third time to be very similar to the second time). This was done to maintain comparability. While the third time is not used for comparison as explained in the supporting information, it can be found in the summary of parameters in Table SI.1. The corresponding bar chart in Fig. 5 shows the fast time constant ($\tau_1$) and the slow time constant ($\tau_2$) in comparison to each other at 290 and 10 K for all measured samples. In the following, we describe our results in a qualitative and not quantitative manner, given the fact that the fast times in the range of a few tens of ps do not represent real decay times, since the temporal resolution of our setup is around 40 to 50 ps. Since no post-processing, such as reconvolution or deconvolution, was performed, all TRPL (substrate-dependent and power-dependent) results remain totally comparable and allow one to extract trends within the uncertainty range of a few ps (of the exponential fitting) at very short time scales.

For all the substrates measured, $\tau_1$ and $\tau_2$ seem to be more pump-density-independent at T=290 K than at T=10 K. This can be explained by noting that biexcitons, which have a higher time constant, are the most density-dependant feature [49]. There are no biexcitons available at T=290 K as shown earlier in Fig. 3 and Tab. 2. At room temperature, WSe$_2$ on SiO$_2$/Si and on MgF$_2$ show longer decay times than the other samples. Additional shortening of the fast decay time is observed as a function of the excitation density. Similarly to Mouri et al. [78], we attribute this to higher exciton-exciton annihilation. Due to the higher pump rates, the necessary diffusion length for exciton-exciton annihilation shrinks. The slow decay time $\tau_2$ shows the same behavior as the fast decay time. Again, the slow decay times for SiO$_2$/Si and MgF$_2$ are longer than the slow decay times for the other substrates. The slow decay times for sapphire (exf.), MgF$_2$ and SiO$_2$/Si get faster with higher excitation powers while the slow decay times for the other two samples remain constant. This increase in decay rates is again attributed to exciton-exciton annihilation. At RT, changes to the temporal characteristics can not only be attributed to the substrate materials properties but also to other effects like interactions with phonons and Auger processes.

At 10 K, the fast decay time at very low pump-densities decreases with increasing power, while at higher pump densities, the decay time remains nearly the same or seemingly increases. A comparison with the power dependence of the different species in the spectrum (Fig. 4) shows that at the lowest power trion, biexciton and localised state emissions equally contribute to the total emission (in the spectrally integrated detection scheme).The exciton does not play a pronounced role. At medium power the fraction of the localised states' emission intensity with respect to the total emission is reduced. At high excitation densities the emission from biexcitons dominates the signal [49].

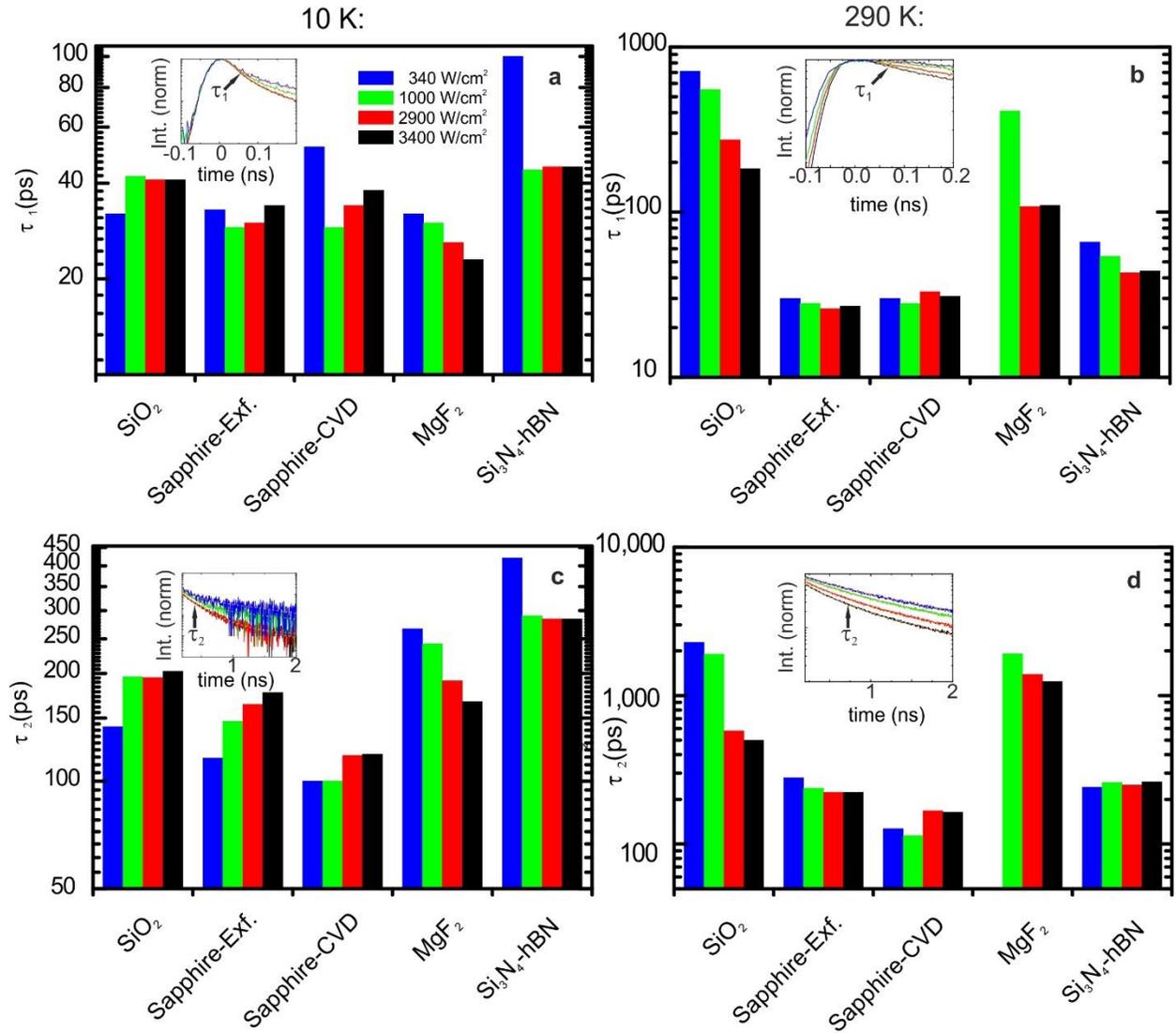

**Figure 5 | Comparison of time constants derived from transient PL of WSE$_2$ monolayers on different substrates at different powers.** The temperature for (a) and (c) is 10K and for (b) and (d) is 290K. The inset in each histogram is an exemplary TRPL plot for WSe$_2$ on SiO$_2$ from which the values of time constant are extracted using triexponential fitting. The overview on all transients is given in Figs. SI.5 and SI.6.

At low temperatures, the substrates become more important as phonon and Auger processes become negligible. Therefore, we revisit the relationship between refractive index and emission properties. Here, a correlation between the refractive index and the general behavior of the fast decay time is indicated. The decay times of the lowest-refractive-index substrate appears to be the fastest while the one for the highest-refractive-index substrate exhibits the slowest decay times. However, this could also reflect the quality of the surface, which at low temperatures could be the significant factor for the exciton dynamics. This reasoning explains the slower decay in the WSe$_2$-hBN/Si$_3$N$_4$ case, where hBN provides an atomically smooth surface for the ML. Simultaneously, spectral properties do not exhibit better excitonic features for this ML-substrate combination, which supports the claim that a non-negligible effect of the dielectric surrounding could be the reason for the fast time constant's trend.

The slow decay times at low temperatures displays two trends. $MgF_2$ and $hBN/Si_3N_4$ exhibit faster decay times with higher pump densities, while the other substrates' decay times slow down slightly with increasing densities. While $MgF_2$ and $hBN/Si_3N_4$ can be fitted with only two time constants, the other samples need to be fitted with three time constants. Nevertheless, all samples have been fitted with three times for better comparability.

All TRPL fit parameters are listed in Table SI.1 of the supporting information. Additionally, Fig. SI.8 in the supporting information summarizes decay times extracted from the TRPL data using a single-exponential fit to the respective data in the range of the second time constant, which is not affected by our setup's resolution and shows similar trends as discussed above.

## Conclusions

We have studied µPL and TRPL from $WSe_2$ deposited on different substrates at both room temperature (290K) and cryogenic temperature (10K). Spectral components such as excitons, trions, biexcitons, and bound states have been identified and compared for different substrates. At room temperature, a small energy shift of the excitonic mode correlating with the refractive index change of the substrate material is indicated. At low temperatures, all ML samples exhibit remarkably similar peak energies for the different species obtained in their emission spectrum. Interestingly, the emission properties of CVD-grown $WSe_2$ on sapphire are very comparable to other ML-substrate cases at low temperature, while at room temperature – in contrast to exfoliated $WSe_2$ on the same material – its emission shows a pronounced red shift of modes, which can be attributed to strain as a consequence of the hot growth process. The relatively high values of the exponent factors α for $WSe_2/SiO_2$, $WSe_2$/sapphire (exf.), and $WSe_2$/sapphire (CVD) at 10K may reflect the dominance of corresponding excitons, trions, and biexcitons among other features in $WSe_2$ PL at cryogenic temperatures. Measured fast and slow decay times of the ML emission, $\tau_1$ and $\tau_2$ at 290K indicate a power-dependent increase of the decay rate which is attributed to exciton-exciton annihilation. Whereas at 10K, the pronounced emergence of excitonic features determines the decay trends with a subtle indication that the refractive index of the dielectric environment may have an effect on the fast decay rates. Thus, this study inspires further detailed investigations concerning substrate-related optical properties of 2D materials and supports the tailoring of application oriented ML systems.

## Methods

**Sample Preparation**

Exfoliation in the group in Marburg: All $WSe_2$ samples are prepared by mechanical exfoliation using bulk $WSe_2$ crystals (*Manchester Nanomaterials*). We exfoliated bulk crystals (*Scotch Magic 3M*) and then transferred them onto a transparent viscoelastic substrate (*Gel-Pak* gel film *PF-30-X4*). Monolayers were then identified by optical contrast using a bright field microscope and transferred onto substrates using the viscoelastic dry-stamping method [79]. All substrates were cleaned in an ultrasonic bath with acetone (99.9% purity) and then rinsed with methanol. $WSe_2$ on $SiO_2$/Si samples were fabricated using a 300nm thermal oxide layer substrate (*IDB Technologies, Ltd*). $WSe_2$ on h-BN/$Si_3N_4$ were fabricated by transferring multilayer h-BN (*Manchester Nanomaterials*) onto a 75nm silicon nitride layer on silicon (*IDB Technologies, Ltd*). $WSe_2$ on $MgF_2$ (*Shanghai OEMC Co., Ltd*) was fabricated using the same technique. After transfer, samples were measured as is and no post processing was applied. All samples were continuously stored in vacuum.

Exfoliation in the group at the Columbia University: Samples were exfoliated onto $SiO_2$/Si (290nm) and sapphire using an established scotch tape mechanical exfoliation method (using *Scotch Magic 3M*). A heated exfoliation was used for TMDC crystals following this, after [80] to increase the size of the monolayers.

CVD growth of $WSe_2$ monolayer on a sapphire substrate: The sample was comprised of a tungsten source carrier chip (5nm $WO_3$ thin film on 90 nm $SiO_2$) and a sapphire substrate (*Ted Pella, Inc.*). Tungsten oxide ($WO_3$, 99.99%, *Kurt J. Lesker*) was deposited on $SiO_2$ via electron beam evaporation. The tungsten source chip was covered, in face-to-face contact, by the sapphire substrate as the growth substrate. The sample was loaded into the center of a 2" diameter and 24" long quartz tube (*MTI Corp.*), and a ceramic boat with 1 g of selenium powder (99.99%, *Sigma-Aldrich*) was located upstream in the quartz tube. After loading, the ambient gas of the tube was purged out via mechanical pump to the base pressure of 10 mTorr. The furnace was heated to 750 ˚C at a 13 ˚C/min ramping rate and hold the temperature at 750 ˚C for 4 minutes, then raise to 850 ˚C at 13 ˚C/min. 20 sccm of Ar gas (5.0 UH purity, *Praxair*) was introduced at 500 ˚C (increasing temperature) to reduce moisture inside of the tube and closed at 500 ˚C (decreasing temperature). Hydrogen (15 sccm, 5.0 UH purity, *Praxair*) gas was supplied to improve $WO_3$ reduction from 700 ˚C (increasing temperature) to 600 ˚C (decreasing temperature). The growth pressure was 1.6 Torr. After 20 min at 850 ˚C, the furnace was cooled down to room temperature naturally.

**Experimental Setup**

μPL: The sample was mounted in a helium-flow cryostat in the μPL setup. All data shown are time-integrated spectra. For excitation, a pulsed titanium-sapphire laser (*SpectraPhysics Tsunami)* with a tuneable emission wavelength of 700-1000nm, a pulse duration 100 fs and a repetition rate of 80 MHz was used. The light from the laser was frequency doubled by nonlinear optics *(CSK Optronics Super Tripler 8315)* to provide an excitation wavelength of 445nm. For detection, a gated intensified charge-coupled device (ICCD) in shutter mode behind a monochromator (*Princeton Instruments Acton SP2300*) was used, using full chip exposure (2D chip read out) and manual integration and truncating of the exposed CCD area. Identification of monolayer positions was performed using extracted integrated spectra which show distinct emission features of monolayers. The image of the sample is focused onto the monochromator entrance slit for spectroscopy. The power-dependent μPL measurements are conducted at 10 K and at 290 K under ultra-high vacuum (~$10^{-6}$ mbar) conditions using a cooling system with liquid helium. Pump-power-dependent measurements were performed at same average power levels, carefully set prior to each measurement run. The investigated powers were set by neutral density filter wheel (discrete steps). The relevant power levels were sequentially changed after the accomplishment of a spectral measurement and its corresponding time-trace measurement. Detection of the µPL signal from the sample takes place behind a spatially filtering aperture in a confocal microscopy geometry. Two long-pass filters were used that block laser light below 650 nm. For evaluation, the recorded 2D spectra (x axis: wavelength, y axis: pixel corresponding to location) is integrated vertically over the relevant pixels corresponding to the excitation spot, minimizing noise and contribution from camera artifacts. The same procedure has been applied to all spectra, integrating over the same amount of pixels. The signal was recorded after optimizing the microscope objective's focus for the most focused projection onto the ICCD (i.e. monochromator entrance slit). With the magnification of the 20x microscope objective and 200 mm focus length of the projection lenses, a reasonable magnification is obtained.

TRPL: Here, highly sensitive fast single-photon counting modules (SPCMs) with timing resolution down to <50 ps (*MPD, PDM Series, Single Photon Avalanche Diodes, 100µm active area*) were used, providing maximum flexibility regarding photon fluxes. The drawback of high count rates is the temporal resolution of the detector. To obtain maximum temporal resolution, a stand-alone Time-Correlated Single Photon Counting (TCSPC) unit with 4 ps histogram time resolution (minimum binning of the *PicoQuant PicoHarp 300*) was employed. Given the instrument response function of the device with decay time between 40 and 50 ps (probed by exposure of the excitation laser), time constants shorter than 40 ps are not considered reliable, even if trends are still seen towards shorter times. No reconvolution or deconvolution has been applied to provide experimental time constants as is, extracted from triexponential fit curves. This is understandable as the measured instrument response using substrate-back-reflected laser light at 445 nm and differing count rates shows a decay of 48 ps slower than the fastest measured sample life times (measured above 700 nm). The detection of the signal from the sample in the µPL setup is similar to µPL spectroscopy, however before the monochromator (using a flip mirror to divert the beam onto the SPCM). Coarse spectral filtering is achieved by long-pass filters, which sufficiently suppress back-scattered laser light and substrate background PL below 650 nm. Before a TRPL measurement is started, monolayer PL is confirmed for the pumped and detected region by time-integrated spectra. No other species than the monolayer PL was detected by the spectrometer for the long-pass filtered signal. Only the $Si_3N_4$ substrate featured a broad background spanning the range of 660 to 820 nm. The $Si_3N_4$ PL background was drastically reduced when an hBN buffer layer was employed. This broad background was brighter than any low-temperature ML signal in case of $WSe_2$ on pure $Si_3N_4$, which forced us to disregard low-temperature PL from $Si_3N_4$. No particular background subtraction has been applied to the spectra other than noise background subtraction, as can be seen in hBN-$Si_3N_4$ spectra. Room-temperature measurements for different sample positions (and flakes) on the same substrate material confirmed strong similarities and reproducibility of emission properties.

**Acknowledgements**

The authors acknowledge financial support by the Philipps-Universität Marburg, the German Research Foundation (DFG: SFB1083) and the German Academic Exchange Service (DAAD). Work at Columbia was funded by the NSF MRSEC under DMR-1420634. AR-I thanks Tony Heinz, Alexey Chernikov, Özgür Burak Aslan for invaluable discussions on 2D materials and properties, and Tineke Stroucken as well as Martin Koch for helpful discussions on excitonic features.

**Author Contributions**

A.R.-I. initiated and guided the joint work on substrate-dependent optical properties of 2D materials in Marburg with the help of Y.D.K., E.H.Y. and J.H. Different sample types were envisioned and fabricated by D.R., D.A., K.K. under supervision of Y.D.K., E.H.Y., J.H. and A.R.-I. The optical setup was established by D.R., M.H., O.A.-M., K.H., S.E.B. and A.R.-I. The experiments were designed and conducted by S.L., L.M.S., D.R., M.H., O.A.-M., X.L., J.K., W. H. and A.R.-I. and the results discussed with the support of all coauthors. The manuscript was written by S.L., L.M.S., K.H., S.E.B. and A.R.-I. with input from all co-authors. S.L. and L.M.S. contributed equally to this work.

## Author Information

The authors declare no competing financial interests.

## Figure legends

**Figure 6 | Microscopic images of the measured WSe$_2$ monolayers deposited on different substrates with cross sections**. The substrates are (a) silicon dioxide, (b) sapphire exfoliated (exf.), (c) sapphire (CVD), (d) MgF$_2$, while in (e), a large flake of multilayer hBN is located underneath the WSe$_2$. Except of (c), the investigated samples have been fabricated by mechanical exfoliation. The yellow line in each image indicates the path along which the brightness cross-sections are taken, which are displayed in the respective charts in the lower row. Given the contrast variations, the layers can be identified and are marked in the cross sections.

**Figure 7 | Schematic diagram of the micro-photoluminescence setup.** The light of the Ti:Sa laser is focused onto the sample using a conventional confocal microscope setup. The setup uses a 20x objective to focus the light onto a 4 μm spot on the sample. The light is collected through the same objective. An iris aperture is used in the sample projection plane for spatial selection of the detection area. The sample can then be imaged using a removable lens and flip mirror in conjunction with a CMOS camera. The light can be focused onto the spectrometer slit for the acquisition of time-integrated spectra with an ICCD or onto an APD for transient μPL measurements, both at room temperature and at 10 K.

**Figure 8 | Time-integrated PL spectra of WSe$_2$ monolayers on different substrates at 290K and 10K for different excitation powers**. The excitation densities shown are 340 (blue), 1000 (green), 2900 (red), and 3400 W/cm² (black). The spectra have been fitted each with the sum of (at least) two Gaussian peaks (light grey dashed lines), with the underlying Gaussian peaks shown in dark grey. For better visibility, the spectra at 10K for the lowest two powers have been scaled.

**Figure 9 | μPL intensity as a function of laser power for different substrates at 290K and 10K**. The power dependence has been plotted on a double-logarithmic scale. The data series correspond to the Gaussian peaks used to fit the PL-spectra. The average peak energies resulting from multi-peak fitting is given in the legend. Furthermore, the series have been fitted to a power law (line) to get further insight into the recombination mechanism and for the sake of comparison. The four data series (symbols) summarized correspond to the features attributed to the exciton (black square), trion (red circle), biexciton (green up-triangle) and localized state (blue down-triangle) in the μPL spectra.

**Figure 10 | Comparison of time constants derived from transient PL of WSE$_2$ monolayers on different substrates at different powers.** The temperature for (a) and (c) is 10K and for (b) and (d) is 290K. The inset in each histogram is an exemplary TRPL plot for WSe$_2$ on SiO$_2$ from which the values of time constant are extracted using triexponential fitting. The overview on all transients is given in Figs. SI.5 and SI.6.

# Influence of the Substrate Material on the Optical Properties of Tungsten Diselenide Monolayers

Sina Lippert[1,+], Lorenz Maximilian Schneider[1,+], Dylan Renaud[1], Kyung Nam Kang[2], Obafunso Ajayi[3], Marc-Uwe Halbich[1], Oday M. Abdulmunem[1], Xing Lin[1], Jan Kuhnert[1], Khaleel Hassoon[1], Saeideh Edalati-Boostan[1], Young Duck Kim[3], Wolfram Heimbrodt[1], Eui-Hyeok Yang[2], James C. Hone[3], and Arash Rahimi-Iman[1,*]

1) Department of Physics and Materials Sciences Center, Philipps-Universität Marburg, Marburg, 35032, Germany

2) Department of Mechanical Engineering, Stevens Institute of Technology, Hoboken, New Jersey, 07030, USA.

3) Department of Mechanical Engineering, Columbia University, New York, New York 10027, USA

[+] these authors have contributed equally

[*] a.r-i@physik.uni-marburg.de

## Supporting Information

**S.1 Raman mode of monolayer $WSe_2$**

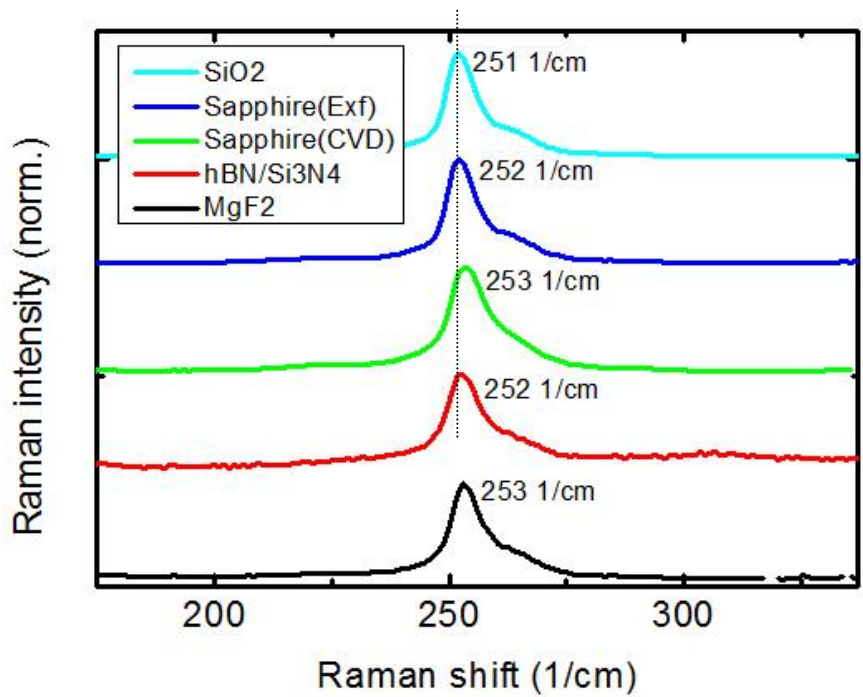

**Figure SI.1 | Raman spectra of ML $WSe_2$ on different substrates measured at 290 K**. The Raman signal detected for the ML deposited on different substrates hardly show a shift in relation to the different substrate materials, with the average wave number of the peaks being 252 cm$^{-1}$. The vertical dotted line is a guide line to eyes.

To further verify that the data was obtained for sample locations which exhibit monolayer (ML) WSe$_2$, room-temperature Raman measurements were performed. For these Raman measurement, a continuous-wave Argon-ion laser with a central wavelength of 514 nm was used. It is coupled to a commercial microscope with 10x, 50x and 100x objective. The reflected light is collected and focused to a spectrometer with a 1200 lines/mm grating. The data is collected by a nitrogen cooled ICCD. An iris aperture is used for spatial selection the same way as in the µ-PL setup, while spectral filtering is achieved with a *holographic super notch filter plus* from *Kaiser optical systems*. One Raman peak can be observed at 250 cm$^{-1}$ with a weak shoulder at 260 cm$^{-1}$. The values of the Raman shift are comparable to the ones given in [S1, S2] Yet, a subtle change in the Raman shift can be observed within the series of samples. This shift might be an indicator for strain [S1]. Additionally, the slight shift for the bilayer WSe$_2$ Raman mode in Ref. [36] under strain to higher wave numbers (for 0% strain at 250 µm$^{-1}$ and for 0.73% at 251 µm$^{-1}$) is another useful indicator of strain for us, since our recorded ML Raman mode for the CVD-grown sample is shifted by more than one wave number when compared to the exfoliated ML on SiO$_2$ (see Fig. SI.1). Moreover, with strain, the shoulder at 260 µm$^{-1}$ disappeared in Ref. [36]. Similarly, our data for the CVD ML does not show a pronounced side peak at higher wave numbers (rather a tail/weak shoulder), while the exfoliated samples show a distinct feature.

## S.2 Spectral properties and fit parameters

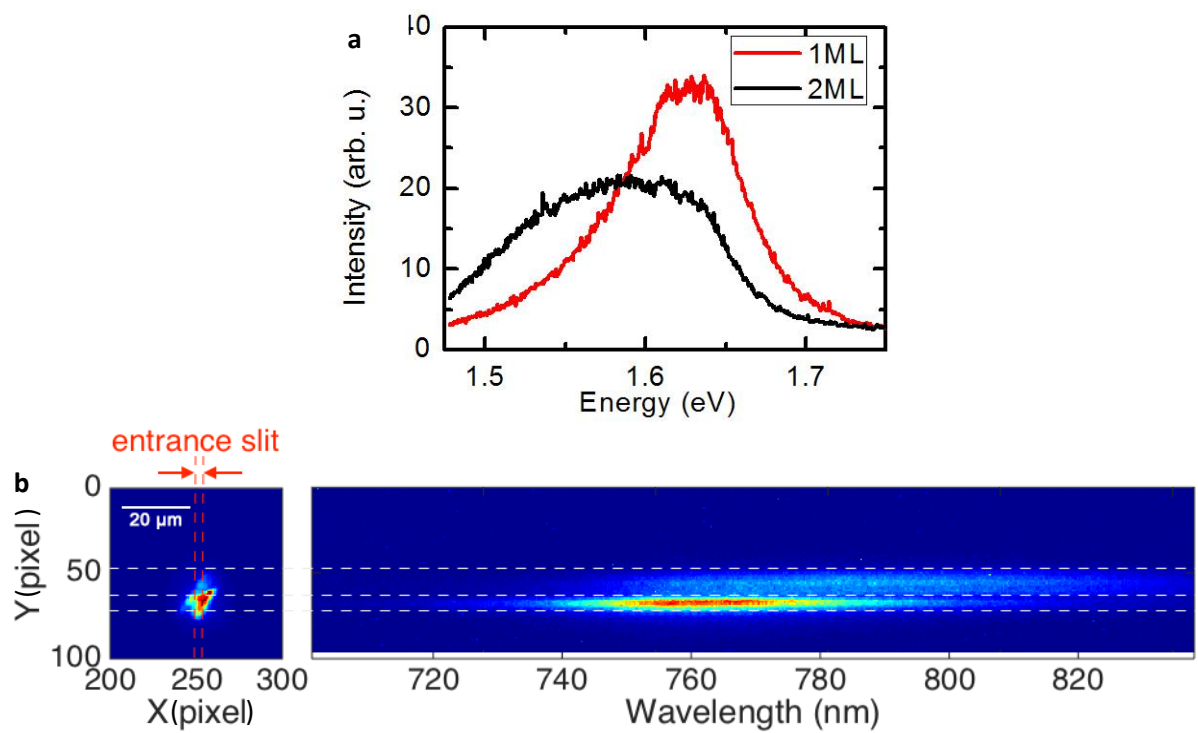

**Figure SI.2 | Monolayer and bilayer emission originating from CVD-grown WSe$_2$ on sapphire.** (a) Comparison of spatially-integrated µPL spectra for the regions of a ML and a bilayer of WSe$_2$. (b) Here, on the left hand side, an image recorded by the monochromator's detector chip shows a micrograph of the emission region, which is pumped with a larger spot than in the usual µPL configuration. On the right hand side, a spatially and spectrally resolved picture is shown, clearly indicating two regions (dashed horizontal lines): one corresponding to a monolayer and another emission region atop of the ML corresponding to a bilayer of WSe$_2$.

Next, a brief view on spatially-resolved µPL is given, which exemplarily was used in combination with a large-area continuous-wave pumping (diameter of pump spot approximately 20 µm) scheme with a 450-nm laser to demonstrate simultaneous excitation and detection of CVD-grown ML and bilayer $WSe_2$ emission on sapphire. This does not represent the actual excitation scheme used in this study, but a variation of it. With the actual pump spot of 4 µm, a very precise excitation of a single ML structure is readily achieved with the µPL setup, providing a strong spatial selectivity. Additionally, a 2D CCD acquisition mode allows for the simultaneous monitoring of spatially differently distributed sample emission. In short, in 2D spectra (as given by the CCD's 2D pixel array), which resolve wavelength on one axis and the sample position along the monochromator's entrance slit on the other axis, emitters can be easily identified and the origin of spectral features along the slit direction observed. Such a scenario is shown in Figure SI.2, in which the identification of ML $WSe_2$ emission is discriminated from bilayer $WSe_2$ signal for CVD-grown flakes on sapphire. In this case, an area was targeted which had two flakes of different contrast vertically aligned, and as expected, the signal from the thinner layer has its peak maximum very close to the expected ML energy position, while the region of higher contrast is spectrally identified as bilayer (2ML) with broad peak centred slightly below 1.6 eV (Ref. [38]), as is shown in Fig. SI.2a. Interestingly, remarkable agreement is given between the bilayer spectrum provided by Ref. [38] at a strain level of "0.73%" (the lowest strain value other than 0% in the measurement series of Desai *et al.* for bilayer $WSe_2$ at room temperature) with our recorded data for a CVD-grown bilayer on sapphire.

In comparison, a micrograph recorded by the monochromator CCD is shown in Fig. SI.2b on the left hand side, while the corresponding 2D spectrum is presented on the right hand side. Here the vertical pixel axis provides spatial information, while the horizontal axis can be transformed to a wavelength scale (as provided by the spectrometer software). Horizontally-dashed white lines mark the two distinct areas corresponding to the ML and 2ML regime, while the entrance slit of the monochromator is indicated by vertically-dashed red lines. In contrast, only a single shiny area (from a ML) was detected in all our spatially-resolved µPL spectra, which allowed us to rule out the occurrence of emission from multilayer $WSe_2$. To obtain line spectra, vertical data integration is performed (along the spatial axis) over the necessary number of pixels which are illuminated by µPL in correlation to the given (strongly focused) pump spot.

In order to attribute spectral components to different emitting species, spectral fit curves were used to get access to single components of a multi peak spectrum. To provide an overview on the reproducibility and comparability of the spectral fit curves, from which the intensities of different species have been extracted, both peak energy positions and line widths are summarized below. The data series shown in Figures SI.3 and SI.4 correspond to Gaussian peaks, which have been used to fit the µPL spectra.

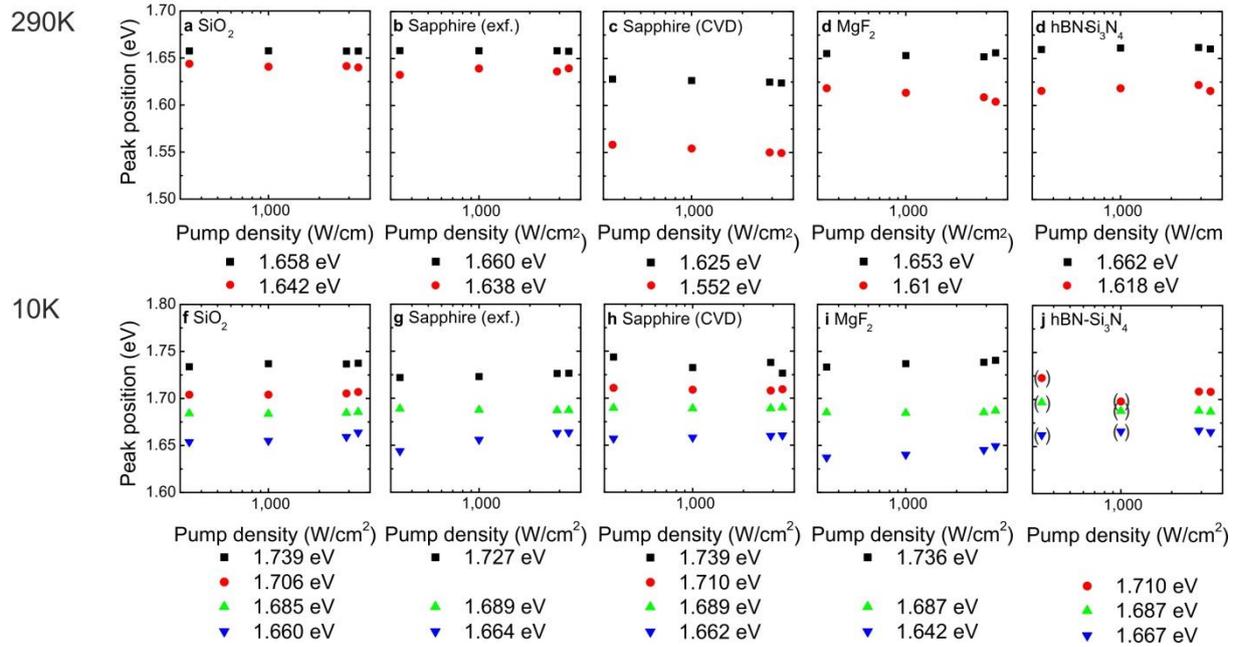

**Figure SI.3 | Peak position as a function of the pump-power density.** For ML WSe$_2$ on different substrates at 290 K and 10 K, the averaged central energy of each species is given in the legend. The peaks' central energies obtained from multi-peak fitting have been plotted in a semi-logarithmic scale. The data series corresponds to the fit curves applied to the µPL spectra (Figure 3) using Gaussian peaks. Data points with high fit uncertainties are indicated by parentheses, given the fact that at low pump densities, the contribution of some species to the signal is hardly distinguishable from the signal noise or background.

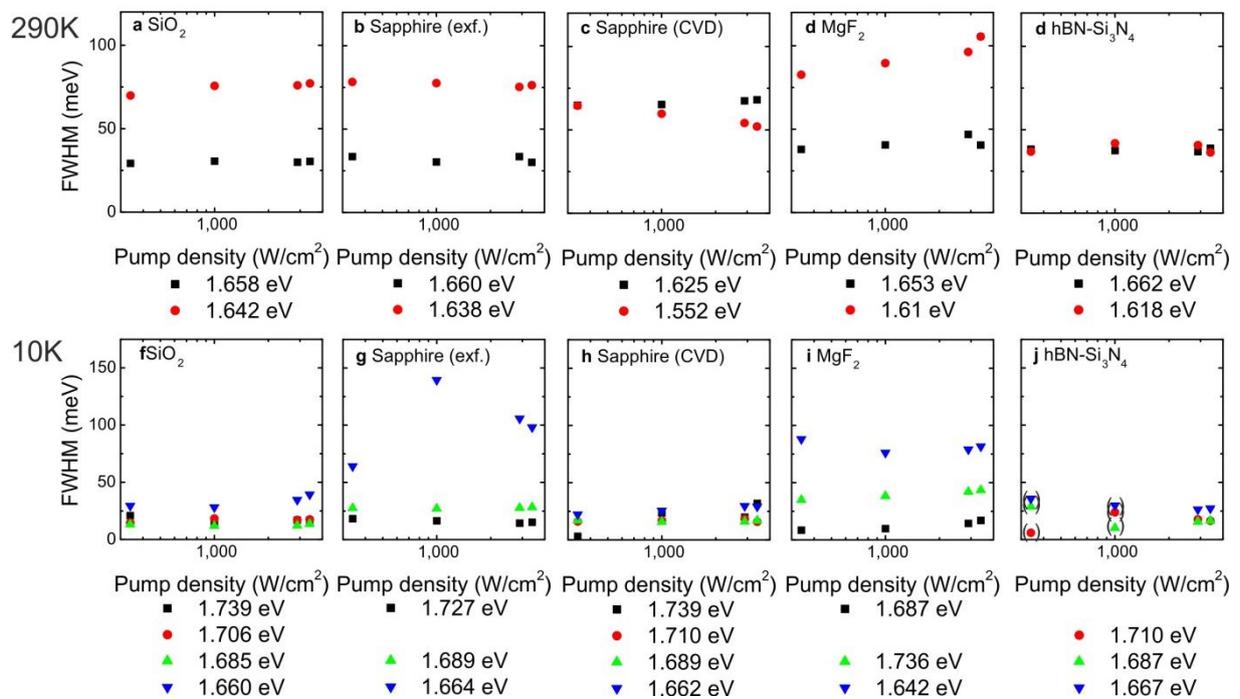

**Figure SI.4 | Full-width half maximum (FWHM) as a function of the pump-power density.** For ML WSe$_2$ on different substrates at 290 K and 10 K, the averaged central energy of each species is given in the legend. The FWHM is one of the parameters used to fit Gaussian multi peaks to the PL spectra. The data is related to the data in Fig. SI.3 and presented in semi-logarithmic scale. Especially for the lowest excitation densities in some cases,

the data was noisy and the signal comparably weak, which caused that fit parameters sometimes exhibit pronounced uncertainty compared to most of the other cases. Such data points are indicated by parentheses.

The peak position as a function of the excitation density is given in Fig. SI.3 for the observed signatures both at 290 K (upper row) and 10 K (lower row). The averaged energetic positions of the peaks are given in the legend. The peak position at room temperature only shows slight changes for the different substrates. The only peak positions which lie at significantly lower energies are those of the CVD-grown $WSe_2$ ML on sapphire. To rule out bilayer emission and to be able to attribute the changes to an effect of strain, we performed a spatially-resolved µPL measurement which covers PL from both a monolayer and a bilayer of CVD-grown $WSe_2$ on sapphire (see above, Fig. SI.2). Remarkably, at low temperature, each species was found at similar spectral positions for the different substrates.

The full-width-half-maximum (FWHM) of the fitted peaks for the investigated samples as a function of the pump-power density is given in Figs. SI.4 both at 290K (upper row) and 10K (lower row). Again, the averaged energetic position of the peaks is given in the legend.

At room temperature, the two Gaussian peaks have the same FWHM in the cases of CVD-grown $WSe_2$ on sapphire (60 meV) and $WSe_2$ on hBN-$Si_3N_4$ (30 meV), respectively. For the other substrates, the FWHM of the peak with higher energy (exciton) is ~30 meV and for the one with lower energy (trion) is ~75 meV. The fact, that all room-temperature-exfoliated MLs' higher-energy peaks exhibit a similar line width at 290 K, indicates that strain in the CVD-grown sample induces line width broadening of the room-temperature excitonic mode (while only slightly affecting the lower-energy mode). In case of the ML deposited on hBN (on $Si_3N_4$), owing to the positive effect of hBN, the line width of the trion is basically restored to the same level of the exciton, being narrower than its counterparts obtained in other ML-substrate combinations.

At low temperature, the FWHM of the different fitted Gaussian peaks of $SiO_2$, sapphire (CVD) and $Si_3N_4$ show similar values (~25 meV). It seems that the species with a particularly small FWHM have excitonic character (e.g. exciton, trion, biexciton, here ~15 meV for nearly all samples which show these features). However, for sapphire (exf.) and $MgF_2$ the FWHM of the second and third peaks is higher in both cases, although to a different extent, correlating to the deviation observed in the extracted α-factors (see Figure 4 of the main text). In both cases, the central peak with energy around 1.688 eV and line width > 25 meV is significantly broader than the exciton line. Especially, the FWHM of the peak attributed to localised states has a significantly higher value. These line width results at 10 K can be seen as an indicator for an increased occurrence of defect states and localized excitons in these two cases, and the suppression of distinct trion and biexciton features, respectively.

With this compilation of parameters from spectral analysis, we can provide a more complete picture of our evaluation and support our results, which demonstrate how similar spectral signatures of ML $WSe_2$ can be for different substrate materials with a rich occurrence of excitonic features at low temperature, and for which ML-substrate combinations significant differences occur.

## S.3 Time-resolved micro-photoluminescence

The time-resolved (TR) micro-photoluminescence (µPL) of the investigated monolayer material (ML $WSe_2$) on different substrate materials was measured first at 290 K (room temperature = RT) and corresponding data is shown in Fig. SI.5 (time-trace histograms). Four different excitation powers were used. The values of the excitation powers were 43 µW (blue), 125 µW (green), 360 µW (red) and 430 µW (black), respectively, using a 100-fs-pulsed 445-nm pump light focused via a microscope objective onto the sample. Given a 4 µm pump spot diameter with Gaussian profile, the mean pump area is estimated with 12.6 µm². Correspondingly, the pump densities amount to approximately 340, 1000, 2900, and 3400 W/cm², respectively. The triexponential fit curves used to extract the decay times are shown as orange dotted lines. The intensities are recorded spectrally integrated and are normalized to the peak intensity for the sake of comparability. Spectral resolution of features for time trace measurements was avoided owing to the lack of signal at these low pump densities and the noticeable variation of spectral features within the series of samples at low temperatures. In other words, since the excitonic features (exciton and charged exciton) at low temperature (10K) exhibit very low signals and are often found as shoulders to more pronounced lower energetic modes (commonly biexcitons and defect states), even with spectral filtering techniques, a comprehensive and comparative analysis was hardly possible. In addition, no absolute comparability of $WSe_2$ signal for different substrate materials is provided as the spectral signatures vary significantly at 10K. Moreover, an SPCM was used for time-trace acquisition by histogram building, which offers higher sensitivity than our Streak camera setup, with which no acquisition was feasible at these pump rates in combination with a monochromator unit attached to it.

The corresponding TR µPL data of the analysed $WSe_2$ monolayers on different substrates measured at 10 K for the same four excitation densities are shown in Fig. SI.6. The same scaling and color code is applied. Again triexponential fit curves, which were used to extract decay times under comparable conditions, are shown in orange. Please note that those time traces which show a late (unrealistic) drop in the intensities are affected by the manual background reduction (in the evaluation process) in case of the noise-floor approaching signal. Furthermore, the signal for transparent $MgF_2$ in the range of the slow decay component was disturbed by a periodically appearing reflection artefact with approximately 400 ps spacing. However, only this room temperature measurement seemed to have suffered from a low count rate and signal quality, which is attributed to the correspondingly unusual spectral behavior (see main text Figs. 3d and 4d) and improper coupling into the SPCM. The lack of a trace at very low excitation densities is explained by the low signal quality in this single measurement series. Nevertheless, the decay times can be still extracted and assessed in comparison to the other cases for the available $MgF_2$-related time traces.

TR µPL was systematically acquired for all samples using the same sequence of measurement steps, that is, before measuring a decay curve, a µPL spectrum at given pump-power density was recorded. Consecutively, after acquisition of µPL and TR µPL for one pump-power setting, the power was changed to the next value. Thus, µPL spectra serve as a means of evidence of ML signal prior to acquisition of time traces by the SPCM.

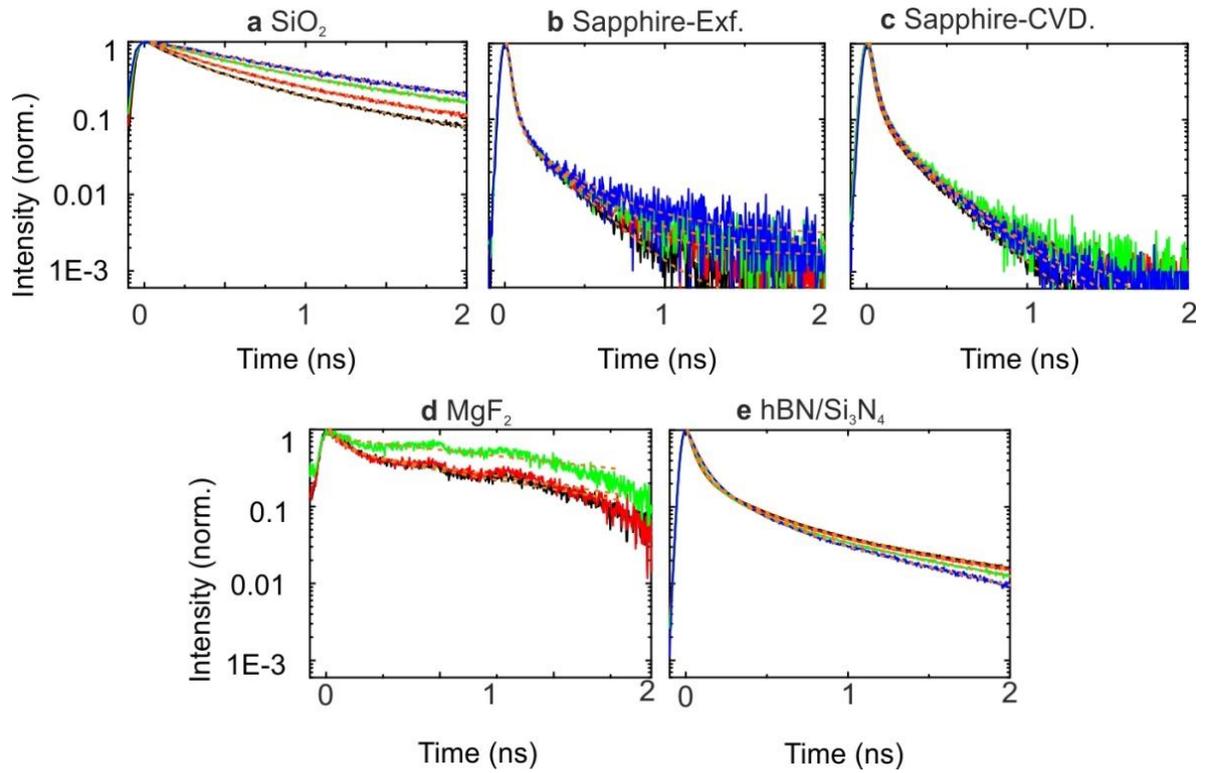

**Figure SI.5 | Transient µPL for different substrates at 290K.** The plots show time-resolved luminescence at four different time-averaged pump-power densities, at 340 W/cm² (blue), 1000 W/cm² (green), 2900 W/cm² (red), and 3400 W/cm² (black), respectively, and as well as fit curves used to extract the decay constants for the considered samples. The scaling and intensity normalization is chosen for the sake of comparability.

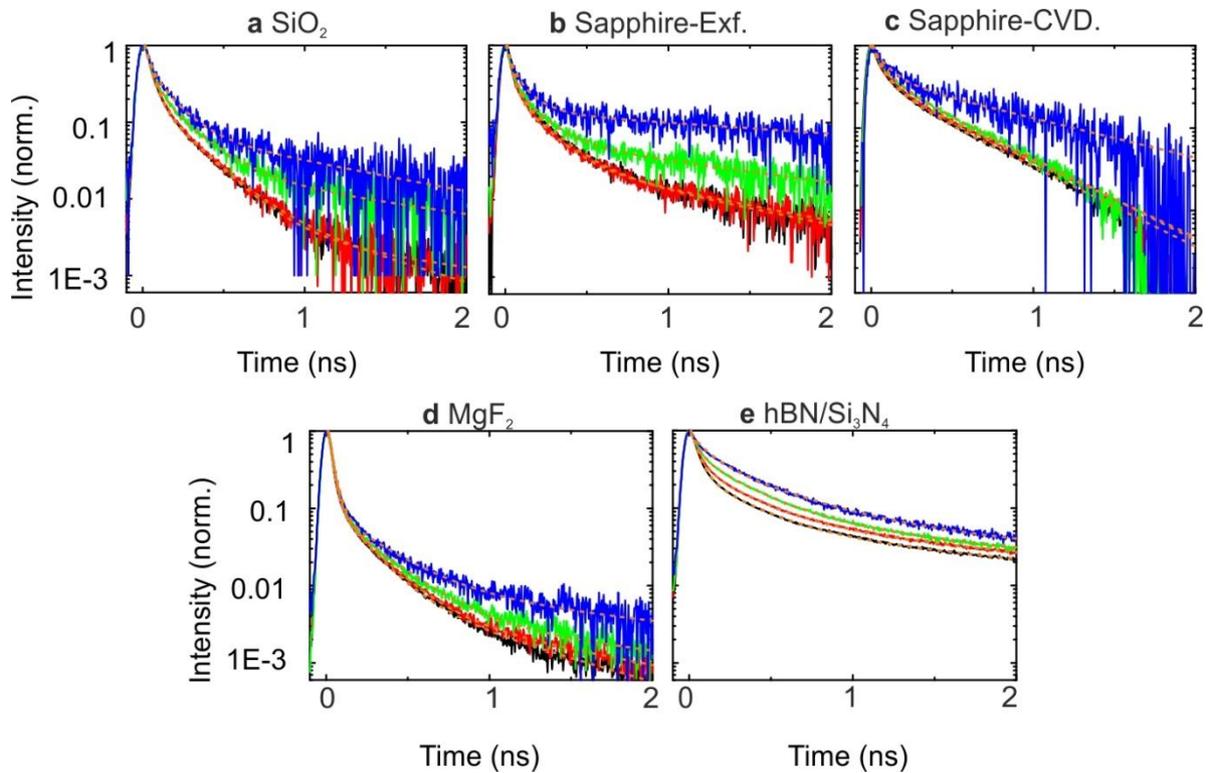

**Figure SI.6 | Transient µPL for different substrates at 10K.** The plots show time-resolved luminescence in a similar fashion as in Fig. SI.5.

## S.4 Decay components of transient PL from WSe$_2$

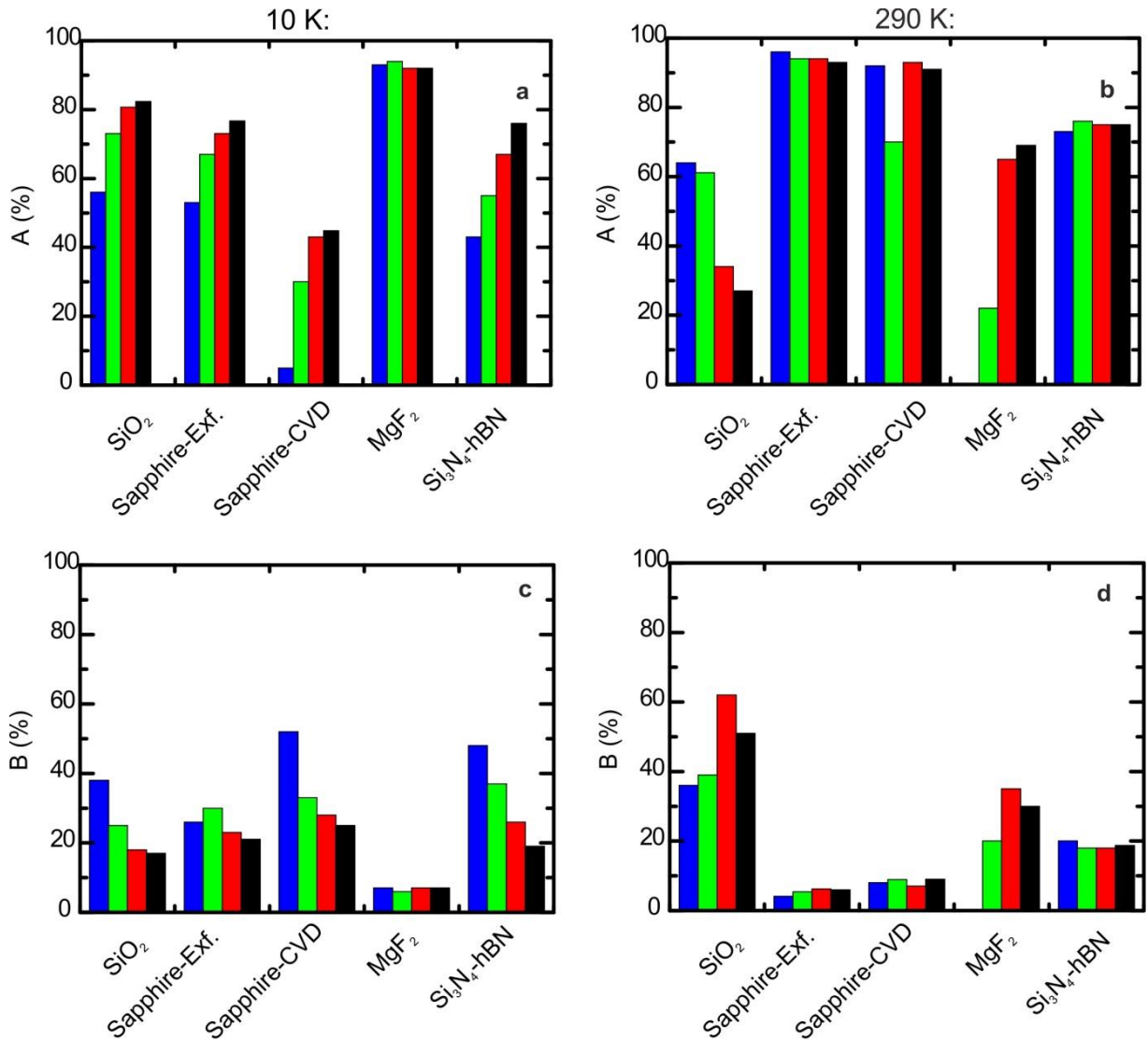

**Figure SI.7 | Relative fit-curve amplitudes A and B of the decay components.** (a & b) fast and (c & d) slow decay of μPL at both 290K and 10K. The color code equals the one of previous figures.

To facilitate consideration of decay constants obtained from TR μPL, the amplitudes of the triexponential fit curves

$$f(t) = A \exp(-t/\tau_1) + B \exp(-t/\tau_2) + C \exp(-t/\tau_3) \qquad (1)$$

are summarized in Fig. SI.7, which shows the amplitude *A* of the fast decay (a and b) and the amplitude *B* of the slow decay (c and d), respectively. The values for these amplitudes represent the fit parameters corresponding to the time traces shown in Figs. SI.5 and SI.6, respectively, and summarized together with the time constants in table SI.1. The different colors represent the four different excitation densities used in the experiments (color code as above). The amplitude is normalized to 100%.

The amplitude *C* of the very slow decay component (on the ns time scale) is comparably low and does not contribute significantly to the signal, with $C = 100\% - A - B$ less than a few percent. Furthermore, in some time traces, the third component cannot be clearly evidenced due to the noise floor and the low intensities, rendering the comparison of this decay component obsolete.

**Table SI.1 | Summary of fit parameters for the measured transients.** The table lists all fit parameters for the triexponential fits. The four values in a row correspond to the four pump densities used, with the densities increasing from left to right. In case the third time constant was similar to the second one, it has been combined with the second one.

|  | SiO2 | Sapphire Exf. | Sapphire CVD | MgF$_2$ | Si$_3$N$_4$/hBN |
|---|---|---|---|---|---|
| **A (10K)** | 0.56/0.73/0.81/0.82 | 0.53/0.67/0.73/0.77 | 0.05/0.30/0.43/0.45 | 0.93/0.94/0.92/0.92 | 0.43/0.55/0.67/0.76 |
| (290K) | 0.64/0.61/0.34/0.27 | 0.96/0.94/0.94/0.93 | 0.92/0.7/0.93/0.91 | -----/0.39/0.50/0.53 | 0.73/0.76/0.75/0.75 |
| $\tau_1$ (10K) | 32/42/41/41 | 33/29/30/34 | 52/29/34/38 | 32/30/26/23 | 100/44/45/45 |
| (290K) | 713/552/274/183 | 30/28/26/27 | 30/28/33/31 | ---/31/81/101 | 66/54/43/44 |
| **B (10K)** | 0.38/0.25/0.18/0.17 | 0.26/0.30/0.23/0.21 | 0.52/0.33/0.28/0.25 | 0.07/0.06/0.07/0.07 | 0.48/0.37/0.26/0.19 |
| (290K) | 0.36/0.39/0.62/0.51 | 0.04/0.05/0.06/0.06 | 0.08/0.09/0.06/0.08 | -----/0.61/0.50/0.47 | 0.20/0.18/0.18/0.19 |
| $\tau_2$ (10K) | 142/196/195/203 | 116/147/164/177 | 100/100/118/119 | 267/242/191/167 | 421/290/284/284 |
| (290K) | 2277/1902/580/500 | 279/238/223/224 | 127/108/167/164 | ---/1911/1388/1242 | 242/259/250/263 |
| **C (10K)** | 0.06/0.02/0.01/0.004 | 0.12/0.03/0.03/0.02 | 0.42/0.37/0.28/0.29 | 0.007/0.003/0.004/0.005 | 0.08/0.07/0.06/0.05 |
| (290K) | -----/-----/0.03/0.21 | 0.003/0.002/1e-3/2e-4 | 3E-4/0.21/0.01/0.01 | ------/-----/------/------ | 0.07/0.006/0.006/0.06 |
| $\tau_3$ (10K) | 1079/1416/710/1494 | 3015/1506/1072/1012 | 903/438/471/448 | 1845/1686/1080/718 | 2673/2034/2060/2123 |
| (290K) | ----/-----/1838/1688 | 4400/6740/6091/603 | 331/399/300/300 | ------/-----/------/------ | 960/1160/1203/1266 |

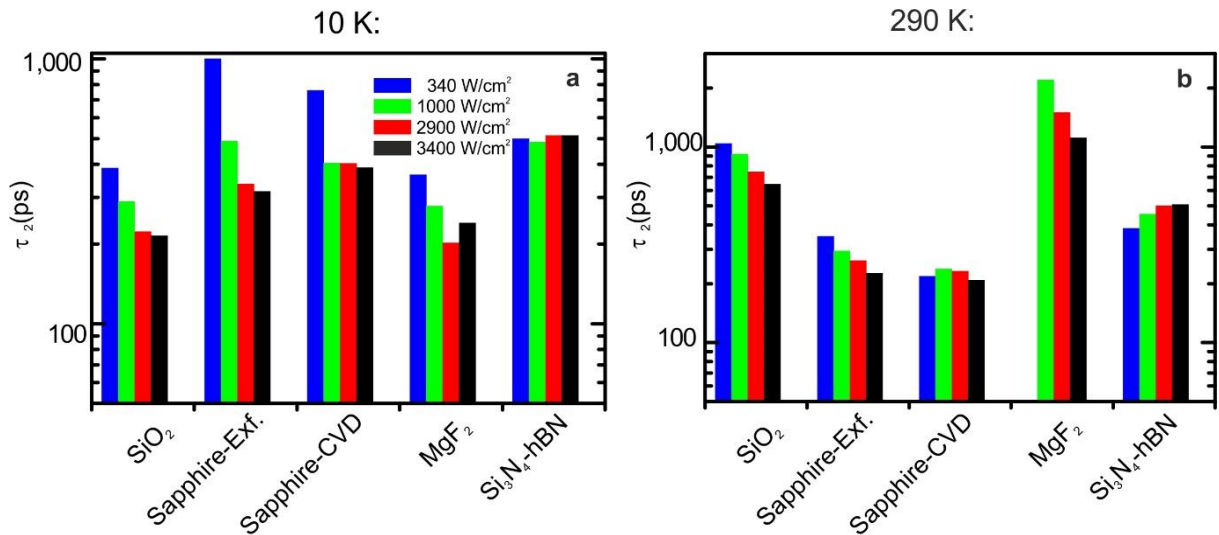

**Figure SI.8 | Second (slow) time constant extracted manually using a single exponential fit.** (a) at 10 K and (b) at 290 K. The color code equals the one of previous figures.

The time constant of fast decay is comparable to the resolution of our whole TRPL setup. We further carried out single exponential fits manually and set the time constant range from 200 ps to 1000 ps which lies in the reliable region of our setup. The fitting results are summarized in Fig. SI.8. The results show a similar trend for RT measurements as in Fig. 5. As to LT measurements, the time constants monotonically decrease as a function of pump density in the studied excitation-density range almost for all samples – most clearly for the MLs on SiO$_2$, sapphire (exf.) and sapphire (CVD). Except for lowest excitation densities, the decay times are on average longest for the hBN-Si$_3$N$_4$, and shortest for SiO$_2$ and MgF$_2$ at LT.

## S.5 Substrate Materials

In Table SI.2, the material details are listed, as provided by the suppliers.

**Table SI.2 | Details on the substrate materials**

| Type | Company Name | Thickness | Roughness | Electric resistance | Crystal plane |
|---|---|---|---|---|---|
| $SiO_2$/Si | IDB TECHNOLOGIES LTD | 300 nm ($SiO_2$) | 0.15-0.3 nm | 1-10 ohm-cm | P<100>B CZ |
| $Si_3N_4$/Si | | 75 nm ($Si_3N_4$) | - | 1-10 ohm-cm | |
| Sapphire CVD | TED PELLA, INC | 1 mm | one microinch or better (polished) | $10^{14}$ ohm-cm @RT | C plane (0001) [S3] |
| Sapphire Exf. | | 0.5 mm | | | |
| $MgF_2$ | Shanghai Optoelectronic Materials And Components Co., Ltd | 1 mm | Ra<0.8 nm [S4] | - | 100 |

## References

[S1] Zhang, R,. Koutsos, V. & Cheung, R, Elastic properties of suspended multilayer $WSe_2$, *Appl. Phys. Lett*. **108**, 042104(2016).

[S2] Sahin, H., Tongay, S., Horzum, S., Fan, W., Zhou, J. and Li, J., Anomalous Raman spectra and thickness-dependent electronic properties of $WSe_2$, *Phys. Rev. B* **87**, 165409 (2013).

[S3] https://www.tedpella.com/vacuum_html/Sapphire_Substrate_Discs_and_Technical_ Information.htm

[S4] http://siiboosh.ce.c-c.com/productinfo/16992245